\documentclass[conference]{IEEEtran}
\IEEEoverridecommandlockouts

\usepackage{amsmath,amssymb,amsfonts}
\usepackage{algorithm}
\usepackage[noend]{algpseudocode}

\usepackage{booktabs}

\usepackage{comment}
\usepackage{caption}

\usepackage{graphicx}

\usepackage{multirow}

\usepackage{pseudocode}

\usepackage{subcaption}
\usepackage{siunitx}

\usepackage{xltabular}

\usepackage{url}

\usepackage{xcolor}
\usepackage{xspace}

\def \bw        {\textsc{bw}}    

\def \CPU       {\textsc{cpu}}
\def \cu        {\textsc{cu}}

\def \dyn       {\textsc{dyn}}

\def \FlowIdx   {\varphi}

\def \idle      {\textsc{idle}}

\def \NameProposal  {GRL-DyP} 
\def \RLBaseline    {Fix-DRL} 
\def \CSP           {CSP}     

\def \NodeIdx       {v}

\def \LinkSet   {L}
\def \LinkIdx   {\ell}

\def \PathIdx   {p}

\def \SFC       {\textsc{sfc}}

\begin{document}

\title{QoS-Aware Dynamic CU Selection in O-RAN with Graph-Based Reinforcement Learning \\
\thanks{This work was supported by NSERC (under project ALLRP 566589-21) and InnovÉÉ (INNOV-R program) through the partnership with Ericsson. We are grateful to Adel Larabi at GAIA, Ericsson Montréal for clarifying some concepts of the current 5G technology.}}

\author{\IEEEauthorblockN{Sebastian Racedo and Brigitte Jaumard}
\IEEEauthorblockA{  \textit{Computer Science and Software Engineering} \\
                    \textit{Concordia University} \\
                    Montreal (Qc) Canada \\
                    brigitte.jaumard@concordia.ca}
\and
\IEEEauthorblockN{Oscar Delgado}
\IEEEauthorblockA{\textit{Systems engineering} \\
\textit{Ecole de Technologie Supérieure (ETS)}\\
Montreal (Qc) Canada}

\and

\IEEEauthorblockN{Meysam Masoudi}
\IEEEauthorblockA{\textit{Ericsson} \\
            Kista, Sweden}
}

%
%

\markboth{Journal of \LaTeX\ Class Files,~Vol.~14, No.~8, August~2015}%
{Shell \MakeLowercase{\textit{et al.}}: Bare Demo of IEEEtran.cls for IEEE Journals}

\maketitle


\begin{abstract}
Open Radio Access Network (O-RAN) dis-aggregates conventional RAN into interoperable components, enabling flexible resource allocation, energy savings, and agile architectural design. In legacy deployments, the binding between logical functions and physical locations is static, which leads to inefficiencies under time-varying traffic and resource conditions. We address this limitation by relaxing the fixed mapping and performing dynamic service function chain (SFC) provisioning with on-the-fly O-CU selection. We formulate the problem as a Markov decision process and solve it using {\NameProposal}, i.e., a graph neural network (GNN)–assisted deep reinforcement learning (DRL). The proposed agent jointly selects routes and the O-CU location (from candidate sites) for each incoming service flow to minimize network energy consumption while satisfying quality-of-service (QoS) constraints. The GNN encodes the instantaneous network topology and resource utilization (e.g., CPU and bandwidth), and the DRL policy learns to balance grade of service, latency, and energy. We perform the evaluation of {\NameProposal} on a data set with 24-hour traffic traces from the city of Montreal, showing that dynamic O-CU selection and routing significantly reduce energy consumption compared to a static mapping baseline, without violating QoS. The results highlight DRL-based SFC provisioning as a practical control primitive for energy-aware, resource-adaptive O-RAN deployments.
\end{abstract}



\section{Introduction}

The transition to 5G and the trajectory toward 6G are accelerating adoption of the O-RAN architecture, shifting networks from proprietary, monolithic stacks to disaggregated, virtualized, and intelligent network \cite{ORAN_2024}. By decoupling the Radio Unit (O-RU), Distributed Unit (O-DU), and Centralized Unit (O-CU), O-RAN enables flexible placement and scaling of functions while fostering a multi-vendor ecosystem. These capabilities are underpinned by Software-Defined Networking (SDN) and Network Function Virtualization (NFV), which also support resource partitioning via network slicing \cite{Kaur_2020}. As the O-RAN Alliance advances specifications and deployment profiles, the resulting design space offers greater agility but also introduces substantial orchestration complexity across heterogeneous hardware, fronthaul constraints, and time-varying traffic \cite{ORAN_2024}.


However, the same flexibility complicates resource allocation and control. Service function chains (SFCs) must be placed, scaled, and steered across heterogeneous compute and transport resources while meeting slice-specific QoS targets, ranging from high-throughput 
enhanced Mobile Broadband (eMBB) to Ultra-Reliable Low Latency Communication (URLLC) \cite{3gpp.23.501}. In practice, many deployments still rely on static deployment and enforce rigid 1:1 bindings among O-RAN components. Such configurations are often derived from offline capacity planning for peak demand, leading to underutilized hardware and significant energy waste during non-peak hours.

The principles of NFV enable resource and routing decisions to be managed by a centralized controller that maintains a global view of the network's state. This opens the door for more intelligent orchestration methods \cite{3gpp.23.501}. 
Although traditional optimization techniques, such as integer linear programs (ILPs), can find optimal solutions, they often struggle to cope with the scale and dynamism of real-world networks \cite{Kaur_2020}. 
This complex, dynamic trade-off space is an ideal application for Deep Reinforcement Learning (DRL). 
A DRL agent, in contrast, can learn complex, non-obvious strategies directly from data patterns, managing the multi-objective problem of maximizing service success while minimizing both latency and energy consumption. 
Besides using DRL, given the natural structure of network-related problems, the use of graph data is usually beneficial.
To work directly with this type of data, we utilize Graph Neural Networks (GNNs) \cite{Wu_2021}.

In this paper, we propose an RL framework that leverages a GNN \cite{Wu_2021} to learn a joint routing and O-CU selection policy. Our agent's architecture is based explicitly on Graph Convolutional Networks (GCNs) \cite{Kipf_2016} that use convolutional network, enabling it to learn from the underlying network topology and real-time state effectively. 
We evaluate our approach, called “{\NameProposal},” using a realistic 24-hour traffic simulation for the city of Montreal. 
To isolate and quantify the benefits of dynamic O-CU selection, we compare its performance to two baselines:
\textit{(i)}    {\RLBaseline} routing-only GNN-GCN DRL policy with fixed, precomputed O-CU placement, and 
\textit{(ii)}   a traditional {\CSP} baseline (Constrained Shortest Path) algorithm that applies a resource constrained shortest-path routing with fixed, precomputed O-CU placement. Our contributions are as follows:
\begin{itemize}
    \item We formulate the joint SFC routing and O-CU selection problem as a Markov Decision Process suitable for a single-agent DRL framework.
    \item We design a GNN-based agent that learns a single, robust policy to handle a full 24-hour traffic cycle for all service types.
    \item We demonstrate through simulation that our agent outperforms a static placement baseline, achieving significant energy savings while respecting strict service-level latency requirements.
\end{itemize}

The remainder of this paper is organized as follows. 
Section \ref{sec:Literature} briefly reviews past work on the provisioning.
Section \ref{sec:Network_description} describes the an augmented version of a realistic synthetic dataset derived from the  city of Montreal (see \cite{jau_CSNM_2023} for the details).
Section \ref{sec:Methodology} presents the proposed methodology. 
Section \ref{sec:Results_Discussion} presents the performance evaluation and discussion of the results, along with their implications.  
Section \ref{sec:Conclusion_FutureWork} concludes the paper and discusses future work.


\section{Literature Review}
\label{sec:Literature}

The increasing complexity of network management, particularly in dynamic environments like O-RAN, has spurred significant research into machine learning (ML) and DRL techniques. While supervised ML offers powerful tools for prediction, DRL has emerged as a key enabler for autonomous decision-making in many areas, one of them being networking \cite{Rusek_2019}. DRL can learn complex control policies directly from experience, optimizing for service-level objectives by maximizing a cumulative reward function \cite{Schulman_2017}. 
This is particularly advantageous for NP-hard problems, such as SFC provisioning, where traditional optimization methods, like ILP, are not able to scale for real-time operations \cite{Jaumard_2024_GNNet}. 
To further enhance a DRL agent's ability to reason about network problems, GNNs have been proposed as a natural fit for modeling network topologies \cite{Wu_2021}, allowing an agent to make more informed, topology-aware decisions.

Several works have applied these advanced learning techniques to the O-RAN and NFV domains. Ali and Jammal \cite{Ali_2024} address the proactive VNF scaling and placement in O-RAN using a multi-stage ML framework. By combining an LSTM traffic forecasting with a DRL agent for placement, their system learns to proactively provision resources, reducing the risk of SLA violations that can occur in reactive systems.

Focusing on the joint SFC embedding and routing problem, Tran \textit{et al.} \cite{Tran_2021} utilize a Deep Q-Learning (DQL) agent. Their agent learns to make sequential node and path selections to provision SFCs under stringent end-to-end delay and bandwidth constraints. Their work demonstrates that a DRL agent can achieve a request acceptance rate of over 95\%, comparable to a traditional heuristic algorithm but with a 10-fold reduction in execution time.

The work by Mei \textit{et al.} \cite{Mei_2021} introduces a hierarchical DRL framework for RAN slicing. Their approach utilizes a high-level controller operating on a large timescale to configure slice parameters (e.g., minimum and maximum data rates) and a low-level controller on a smaller timescale to perform real-time resource scheduling. 
This demonstrates the power of hierarchical control for managing problems with different temporal granularities. 
Similarly, Mollahasani \textit{et al.} in \cite{Mollahasani_2021} use a nested actor-critic model where a high-level agent decides on the optimal placement of a network function (O-DU or O-CU) to balance observability and latency, while a low-level agent handles the specific resource block allocation.

Addressing higher-level orchestration, Staffolani \textit{et al.} \cite{Staffolani_2024} introduce PRORL, a DRL agent that learns to allocate resource units from a central pool to various Points of Presence (PoPs). By framing the problem as a Multi-Objective MDP, their agent learns to balance demand satisfaction, resource utilization, and the operational cost of moving resources, achieving an average improvement of 90\% over greedy baselines.

To address the inherent graph structure of computer networks, GNNs have been proposed to enhance learning capabilities. 
The work by Rusek \textit{et al.} in \cite{Rusek_2019} specifically highlights this potential with their GNN-based model, RouteNet. They demonstrate that a GNN trained on one network topology can successfully generalize to make accurate performance predictions on larger, unseen topologies, a critical capability for real-world networking applications.

While these works establish a strong foundation, several research gaps remain. A common limitation is the choice of topology: solutions are often tested on small networks that seldom exceed a dozen nodes. 
Secondly, the types of constraints are not always consistent across the board: some solutions consider delay limits for requests without including VNF processing requirements, while others focus on resource allocation without per-flow routing decisions. 
To address these gaps, our work considers a larger, more realistic network topology and integrates multiple constraints, including latency, bandwidth, CPU capacity, and, critically, the energy consumption of both network nodes and function processing, an aspect often overlooked in prior SFC provisioning studies.


\section{Network Topology and Problem Statement}
\label{sec:Network_description}


\subsection{Network Topology}

    The physical network is represented as a directed graph $G=(V, E)$, where $V$ represents the set of physical nodes and 
    $E$ is the set of fiber or microwave links. 
    A subset of these nodes hosts compute resources and can run multiple co-located VNFs (e.g., a O-DU and a O-CU). 
    This subset of nodes is heterogeneous and has bounded CPU, memory and storage resources. 
    A bandwidth capacity and a propagation delay characterize each bidirectional link $\LinkIdx \in \LinkSet$. 
    Our scenario includes RUs distributed across the region. 
    These RUs handle both uplink and downlink traffic flows. 

    
    Each flow is defined by its service type, source/destination, required bandwidth, and a maximum end-to-end latency constraint. More details on each service SFC can be found in Table \ref{tab:SFC_ziazet_2023} and in \cite{Ziazet_2023_FNFW}.
    
    \begin{table}[ht]
        
        \caption{SFC (values adapted from \cite{Ziazet_2023_FNFW})}
        \resizebox{\columnwidth}{!}{%
        \centering
        \begin{tabular}{c|c c c c c}
            \hline
            Service  &  Class  & SFC  & Bandwidth  & Latency  &  \# Request\\ 
                       
            \hline \hline
         Cloud  & \multirow{2}{*}{eMBB}    &  \multirow{2}{*}{UPF$_{\textsc{CG}}$- CU - DU - RU} & \multirow{2}{*}{4 Mbps} 
                & \multirow{2}{*}{80 ms} & \multirow{2}{*}{[40-55]} \\ 
             gaming (\textsc{CG}) & & & & & \\
             \hline
             Augmented  & \multirow{2}{*}{eMBB}    & \multirow{2}{*}{UPF$_{\textsc{AR}}$- CU - DU - RU} & \multirow{2}{*}{100 Mbps} 
                        & \multirow{2}{*}{20 ms} & \multirow{2}{*}{[1-4]} \\ 
             reality (\textsc{AR}) & & & & & \\
             \hline
             \multirow{2}{*}{VoIP}  & \multirow{2}{*}{eMBB}    & RU - DU - CU - UPF$_{\textsc{VoIP}}$ & \multirow{2}{*}{64 Kbps} 
                                    & \multirow{2}{*}{100 ms} & \multirow{2}{*}{[100-200]} \\ 
             & &  UPF$_{\textsc{VoIP}}$- CU - DU - RU & & &  \\
             \hline
             Video  & \multirow{2}{*}{eMBB}     & \multirow{2}{*}{UPF$_{\textsc{VS}}$- CU - DU - RU} 
                    & \multirow{2}{*}{4 Mbps}   & \multirow{2}{*}{100 ms}       & \multirow{2}{*}{[50-100]} \\ 
             streaming (\textsc{VS}) & & & & & \\
             \hline
             Massive    & \multirow{2}{*}{mMTC}    & \multirow{2}{*}{RU - DU - CU - UPF$_{\textsc{MIoT}}$} & \multirow{2}{*}{[1 -50] Mbps} 
                        & \multirow{2}{*}{10 ms} & \multirow{2}{*}{[10-15]} \\ 
             IoT (\textsc{MIoT}) & & & & & \\
             \hline
             Industry 4.0   & \multirow{2}{*}{uRLLC}    & RU - DU - CU - UPF$_{\textsc{I4.0}}$ 
                            & \multirow{2}{*}{70 Mbps} & \multirow{2}{*}{15 ms} & \multirow{2}{*}{[1-4]} \\
             (I4.0)         & & UPF$_{\textsc{I4.0}}$- CU - DU - RU &  &  & \\
             \hline
        \end{tabular}
        } 
        \label{tab:SFC_ziazet_2023}
    \end{table}

\subsection{Problem Statement}
    Given the network topology, the set of possible locations for each VNF type, and the 24-hour traffic, the problem is to provision each incoming SFC request dynamically. This involves two joint decisions:
    \begin{enumerate}
        \item O-CU Selection: Selecting a suitable O-CU from the set of pre-deployed candidate nodes to host the flow's O-CU function.
        \item Routing: Determining a physical path from the flow's source to its destination while satisfying all the constraints.
    \end{enumerate}

    A provisioned SFC is considered valid only if it satisfies all of the following constraints:
    Let $\PathIdx$ be the selected end-to-end path for a flow, consisting of a sequence of links $\LinkIdx \in \PathIdx$. 
    Let $V_{\SFC} \subset V$ be the set of nodes where the flow's VNFs are placed.
    
    \begin{itemize}
        \item \textbf{CPU Capacity:} For the selected O-CU node, $v_{\cu} \in V_{\SFC}$, its available CPU resources must be sufficient to handle the demand generated by the flow $\FlowIdx$:
        \begin{equation}
            {\CPU}_{\text{free}}(v_{\cu}) \geq Demand_{\CPU}(\FlowIdx).
        \end{equation}
        \item \textbf{Bandwidth Capacity:} For every link $\LinkIdx$ in the chosen path $\PathIdx$, the available bandwidth must be greater than or equal to the flow's requirement:
        \begin{equation}
            {\bw}_{\text{free}}(\LinkIdx) \geq {\bw}_{req}(\FlowIdx) \qquad \LinkIdx \in \PathIdx.
        \end{equation}
        \item \textbf{QoS Constraint:} The total end-to-end delay, comprising the sum of propagation delays ($\delta_{prop}(\ell)$) on the links and the processing delays at each VNF node ($\delta_{proc}(\NodeIdx, \FlowIdx)$), must not exceed the flow's maximum allowed latency ($L_{\max}(\FlowIdx)$):
        \begin{equation}
            \sum\limits_{\LinkIdx \in \PathIdx} \delta_{prop}(\LinkIdx) + 
            \sum\limits_{\NodeIdx \in V_{\SFC}} \delta_{proc}(\NodeIdx, \FlowIdx) \leq L_{\max}(\FlowIdx).
        \end{equation}
    \end{itemize}

    The overall objective is to learn a policy $\pi$ that, for each incoming flow, finds a valid provisioning solution that is optimal with respect to our defined reward function. The goal is not to find all possible valid paths, but to find the one that best maximizes the cumulative reward, which encapsulates the trade-off between successfully granting the flow and minimizing the energy cost of the chosen path and O-CU selection. This implicitly maximizes the number of successfully provisioned flows over the 24-hour period by using resources efficiently.


\subsection{Power Model}

    To evaluate the agent's performance, we model the network's power consumption. 
    For the processing nodes (O-DU, O-CU, UPF), we use the classical and widely used linear power model ($P_{node}$) proposed by Fan \textit{et al.} \cite{Fan_2007} which relates CPU utilization ($u_{node}^{\CPU}$) to power draw.
    For transport nodes acting as routers, we use \cite{Vishwanath_2014}, which is based on idle power ($P_{net}^{\idle}$) and per-bit forwarding energy ($P_{net}^{dyn}$).
    \begin{alignat}{2}
        & P_{node}          && = P_{node}^{\idle} + (P_{node}^{\max} - P_{node}^{\idle}) \times u_{node}^{\CPU} \\
        & P_{net}^{\idle}   && = \sum_{r \in R} P_{r}^{\idle} + \sum_{\LinkIdx \in \LinkSet} P_{\ell}^{\idle} \\ 
        & P_{net}^{\dyn}    && = \sum_{r \in R} \left( \frac{P_{r}^p}{8L} + \frac{P_{r}^{\textsc{sf}}}{8}\right),
    \end{alignat} 
    where 
    $R$ is the set of routers, 
    $\LinkSet$ is the set of links, 
    $P^p$ is per-packet processing power, and 
    $P^{\textsc{sf}}$ is per-byte store and forward power.
    

\section{Methodology}
\label{sec:Methodology}

To address the dynamic SFC provisioning challenge, we model the problem as a MDP and simulate an RL environment where the agent sequentially provisions one flow per episode, we assume that the resources of each flow remain in the network per hour, meaning that per hour we reset the overall state of the network. The environment exposes the network graph, resource states, and traffic requests sampled from the 24-h trace, enabling the agent to interact with a dynamic simulator. The agent is trained using Maskable PPO \cite{Huang_2022}, an extension of PPO \cite{Schulman_2017} that handles dynamic action spaces.

    
\subsection{MDP Formulation}

The state $s_t \in S$ comprises 
\textit{(i)} a task vector with flow metadata and 
\textit{(ii)} node features encoding topology and resources. 
The environment also provides an action mask at each step to nullify invalid actions. Details on these components are as follows:
        \textit{(i)} \textbf{Task Vector (`obs`):} A low-dimensional vector with flow-specific information: the service ID, the current hour of the day, the current SFC segment index, the agent's current location index, the target location index for the current segment, the graph distance to the target, and the remaining latency budget. 
        \textit{(ii)} \textbf{Graph Features (`node\_features`):} An $N \times F$ matrix describing the network, where $N$ is the number of nodes and $F$ is the number of features per node. These features are done per node and include: a multi-hot encoding—a binary vector where multiple entries can be active, indicating all logical functions a node can host (e.g., O-DU, O-CU, UPF)—, the normalized number of connections the node has to other nodes, and the real-time available CPU of the node (if it has compute capacity), which provides a direct signal for network congestion. 

        We define a joint action space $A$ for the agent, allowing for the joint decision of routing and embedding. The action space is a discrete set of size $D_{\max} + 1$, where $D_{\max}$ is the maximum out-degree of any node in the graph.
        \begin{itemize}
            \item \textbf{Routing Actions ($a \in [0, D_{\max} - 1]$):} Correspond to a `\textsc{move}` action, where the agent traverses a link to a neighboring node.
            \item \textbf{Embedding Action ($a = D_{\max}$):} Corresponds to an `\textsc{embed}` action. This action is only available during the O-DU-to-O-CU routing segment and allows the agent to select its current location for the O-CU function.
        \end{itemize}

The action mask (`action\_mask`) is a binary vector indicating which actions are valid in the current state, $s_t$. This mask is generated by the environment at each step to prevent the agent from selecting actions that would violate hard constraints.


The task vector has 7 dimensions; node features form an $N \times F$ matrix ($F$=5 per node: roles, degree, CPU, etc.). $D_{\max}$ is the maximum node out-degree (8 in our topology), so the joint action space size is $D_{\max} + 1$. Routing is performed per-flow, step by step, until the SFC is provisioned.

        To guide the agent towards the dual objectives of QoS satisfaction and energy efficiency, the total reward for an episode is a summation of components awarded at each step. The agent's objective is to maximize the cumulative discounted reward, where the reward at each step $RW(s_t, a_t)$ is composed of several event-driven components. A conceptual formula for the total reward of a complete trajectory $\tau$ is:
        \begin{multline} 
        \label{eq:reward}
        RW_{\text{total}}(\tau) = RW_{terminal} + \sum_{t \in \tau} (rw_{\text{shaping}} \\
        + rw_{\text{intermediate}} - p_{\text{energy}}).
        \end{multline}
        Each term of the reward function is defined as follows:
        \begin{itemize}
            \item \textbf{Dense Shaping Reward ($rw_{\text{shaping}}$):} At every `MOVE` action, the agent receives a small, immediate reward proportional to the reduction in shortest-path distance to its next target. It also receives a penalty for creating loops by revisiting nodes.
            
            \item \textbf{Intermediate Success Reward ($rw_{\text{intermediate}}$):} Upon completing a segment (either by arriving at a pre-defined VNF or by performing a valid `EMBED` action), the agent receives a large positive reward. This reward is penalized by the number of hops taken within that segment to encourage path efficiency.
            
            \item \textbf{Energy Penalty ($p_{\text{energy}}$):} The intermediate reward for a successful `EMBED` action is penalized based on the estimated power consumption of the chosen node. Furthermore, the final success reward is penalized by the total energy consumed by all nodes in the chosen path. This directly guides the agent to find energy-efficient solutions.
            
            \item \textbf{Terminal Reward ($RW_{\text{terminal}}$):} At the end of the episode, a significant positive reward is given for successfully provisioning the entire SFC. A significant negative penalty is given for any action that leads to a terminal state by violating a constraint (e.g., exceeding the latency budget or reaching a dead end).
        \end{itemize}

    \subsection{GCN Architecture}
    Our GCN architecture comprises: \textbf{GCN Backbone:} Three GCNConv layers that process the `node\_features` matrix and the graph's `edge\_index` 
    to produce a rich embedding for each node in the network. \textbf{Feature Embedding:} Separate embedding layers for the categorical features in the `obs` vector (service ID, hour, function type, etc). \textbf{MLP Heads:} The GNN embedding of the agent's current node is concatenated with the embedded features and fed into a shared MLP. The output of this MLP is then passed to two separate linear layers to produce the policy logits (for the Actor) and the state-value prediction (for the Critic).


\section{Results \& Discussion}
\label{sec:Results_Discussion}

We now discuss the validation of the proposed {\NameProposal} models and algorithms.


\subsection{Energy Savings of {\NameProposal}}

\begin{figure*}[ht]
        \centering
        \includegraphics[width=1.0\textwidth]{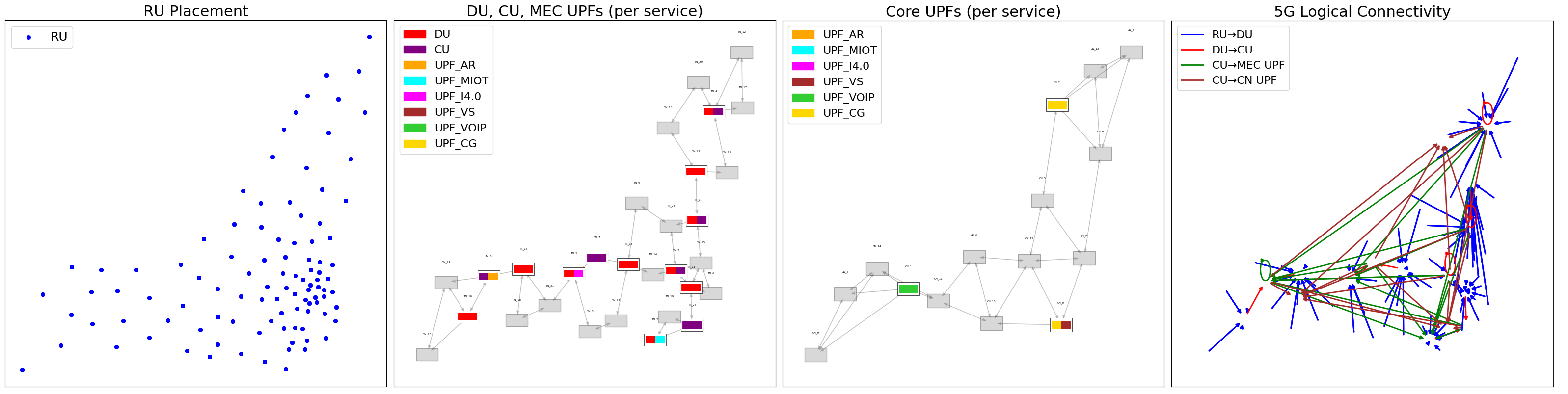}
        \caption{Fixed placement of logical functions capable of accommodating a dynamic traffic over 24 hours}
        \label{fig:placement_result_pcenter}
    \end{figure*}

\begin{figure}[ht]
        \centering
        \includegraphics[width=0.75\linewidth]{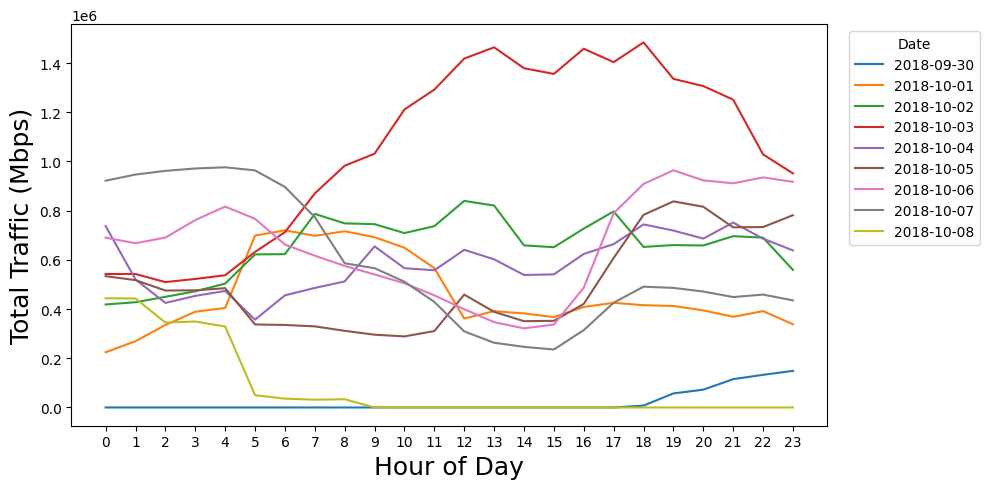}
        \caption{Overall traffic per hour, by day}
        \label{fig:overall_traffic}
    \end{figure}

\begin{figure*}[ht]
    \centering
        \includegraphics[width=0.7\textwidth]{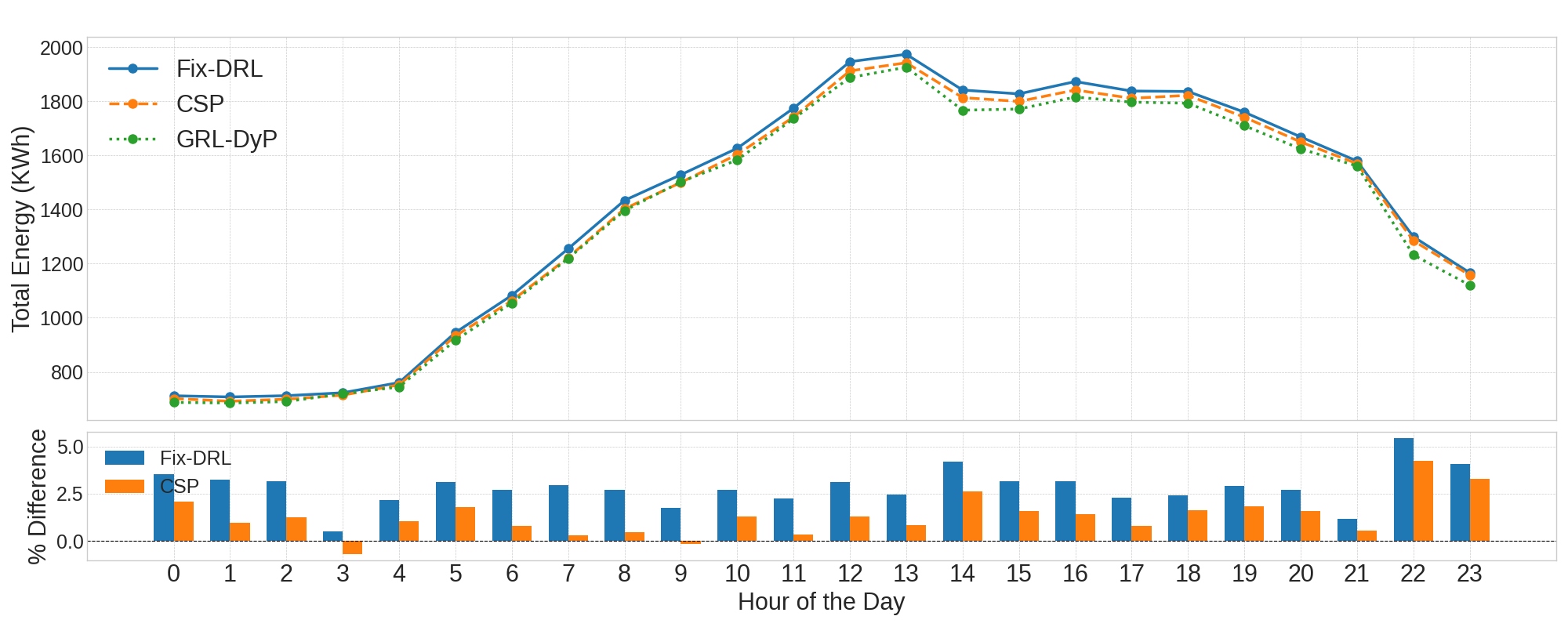}
        \caption{Total energy consumption per hour comparison}
        \label{fig:energy_hourly_comparison}
    \end{figure*}

\begin{figure}[ht]
        \centering
        \includegraphics[width=0.9\columnwidth]{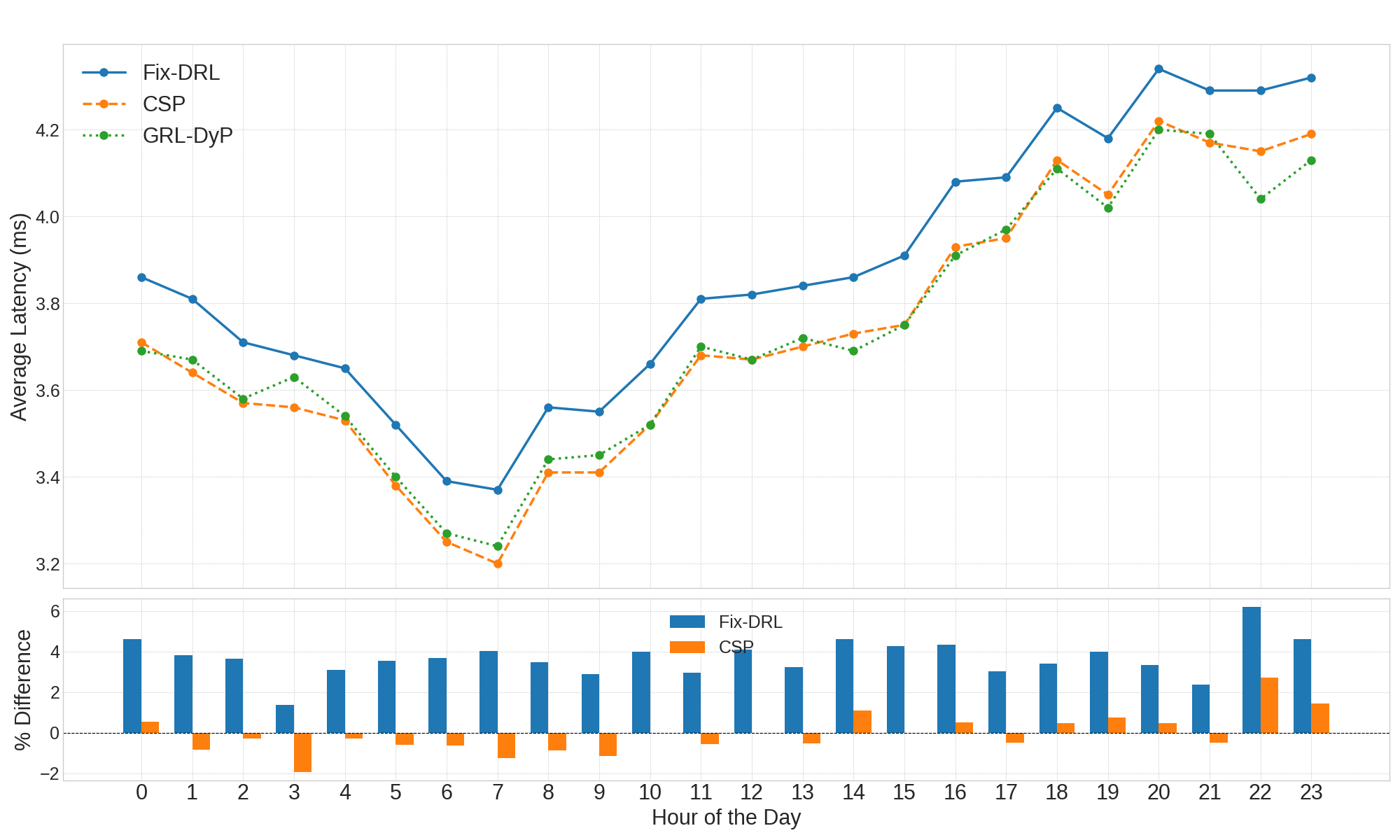}
        \caption{Average latency per hour comparison}
        \label{fig:latency_hourly_comparison}
    \end{figure}

    Our scenario consist of 300 O-RUs, 10 O-DUs, 6 O-CUs and 7 UPFs that are distributed across a 5G network. 
    Figure \ref{fig:placement_result_pcenter} illustrates the initial placement of the logical functions used in our scenario.
    To do the placement, we first analyzed the daily traffic generated by all service slices, as shown in Figure \ref{fig:overall_traffic}. From this analysis, we identified the 24-hour period with the highest total traffic. This peak-demand data was then used to determine the placement of the logical functions (in our scenario each slice corresponding to a single service type). Traffic traces in Figure \ref{fig:overall_traffic} are the raw dataset; training samples flows from this day, while evaluation uses disjoint subsets.
    We evaluate two DRL agents: (1) {\RLBaseline}, which learns a routing-only policy for a fixed 1:1 O-DU-to-O-CU mapping, and (2) {\NameProposal}, which learns the joint routing and dynamic O-CU selection policy (both RL agents were trained for 1M steps using Maskable PPO from  Stable-Baselines3 contrib, with roll-out buffer 4096, batch 128, lr=$3e^{-4}$, $\gamma=0.99$, clip=0.2, and GCN hidden dim=64. Training used sampled flows; testing used a separate 24-hour traffic dataset with similar distribution patterns, ensuring evaluation under comparable but unseen traffic conditions. Figures \ref{fig:energy_hourly_comparison}–\ref{fig:latency_hourly_comparison} report the total energy and latency across the entire subset, not a single request, with flows provisioned dynamically to emulate online arrivals). We compare them against a {\CSP} algorithm that uses the same fixed 1:1 mapping and routes flows via the shortest delay valid path. The simulation uses a realistic 24-hour traffic trace from the city of Montreal, see \cite{jau_CSNM_2023} for the details. 
    Our primary baseline is a heuristic that reflects how a network operator might eagerly route traffic, with all flows served with minimal delay.
    \begin{figure}[!t]
        \centering
        \includegraphics[width=0.9\columnwidth]{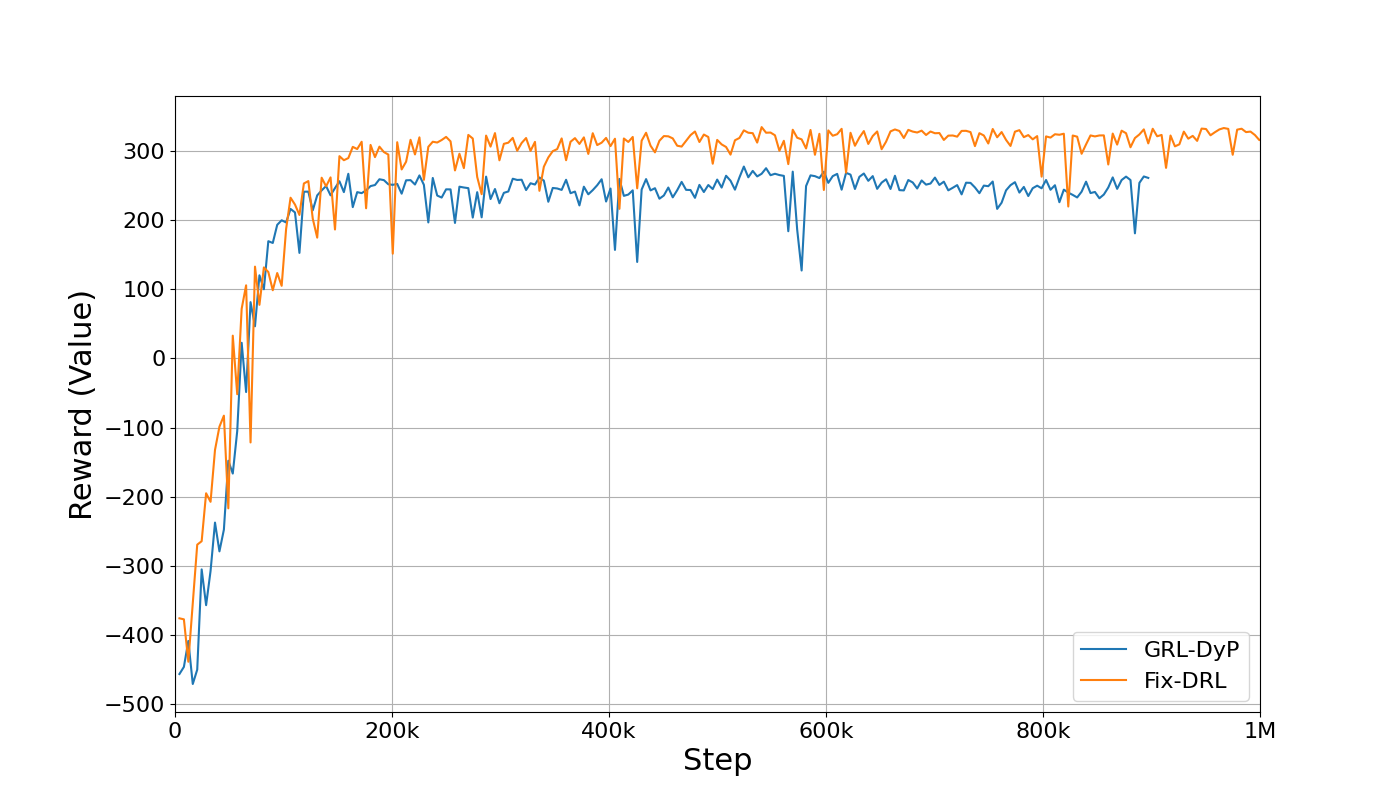}
        \caption{Learning curves for the {\NameProposal} and {\RLBaseline}}
        \label{fig:training_reward_curves}
    \end{figure}
    Figure \ref{fig:training_reward_curves} shows the learning curves for our two DRL agents. Both agents exhibit stable learning, with the average episode reward converging to a high positive value. This demonstrates that the Maskable PPO algorithm, guided by our reward structure, is capable of solving the complex routing and embedding tasks.
    
    The primary results of our study are shown in Figures \ref{fig:energy_hourly_comparison} and \ref{fig:latency_hourly_comparison}. 
    The {\NameProposal} policy identifies O-CU nodes where embedding can be performed, reducing total energy consumption during evaluation by
    1.3\% (447.8 kWh) vs. {\CSP} and 2.85\% (1,039.5 kWh) vs. {\RLBaseline} over 24 h, while still meeting all demand and latency constraints. 
    
    
    Furthermore, the {\NameProposal} is able to learn a routing strategy that is more congestion-aware than the simple shortest-path with constraints baseline. As seen in Figure \ref{fig:latency_hourly_comparison}, the agent maintains a consistently similar or lower average latency, particularly during the peak traffic hours.

    \begin{table*}[ht]
    \centering
    \caption{Combined Per-Slice Performance Metrics}
    \label{tab:combined_metrics_latency}
    \resizebox{0.70\textwidth}{!}{%
    \begin{tabular}{@{}lcccccc@{}}
        \toprule
        \textbf{Slice} & \multicolumn{2}{c}{\textbf{\RLBaseline}} & \multicolumn{2}{c}{\textbf{\CSP}} & \multicolumn{2}{c}{\textbf{\NameProposal}} \\
        \cmidrule(lr){2-3} \cmidrule(lr){4-5} \cmidrule(lr){6-7}
        & \textbf{Avg. Latency (ms)} & \textbf{Avg. Links Used} & \textbf{Avg. Latency (ms)} & \textbf{Avg. Links Used} & \textbf{Avg. Latency (ms)} & \textbf{Avg. Links Used} \\ \midrule
        Augmented reality  & 6.76 & 8.27 & 5.48 & 8.02 & 6.82 & 8.62 \\
        Cloud Gaming       & 4.41 & 7.08 & 3.45 & 6.81 & 4.46 & 7.31 \\
        Industrie 4.0      & 3.74 & 6.36 & 3.41 & 5.73 & 3.75 & 6.39 \\
        Massive IoT        & 4.86 & 9.67 & 4.35 & 8.36 & 4.68 & 8.79 \\
        Video streaming    & 4.51 & 6.64 & 4.10 & 5.99 & 4.47 & 6.49 \\
        VoIP               & 2.17 & 7.26 & 1.97 & 6.46 & 2.05 & 6.63 \\ \bottomrule
    \end{tabular}%
    }
    \end{table*}
    
    The proposed policy \NameProposal achieves significant energy savings without compromising service quality, as confirmed by all six service slices meeting their latency requirements (Table \ref{tab:combined_metrics_latency})


    \subsection{Discussion}

    The results highlight a clear advantage for dynamic, AI-driven orchestration. The primary source of energy efficiency is the agent's learned ability to perform dynamic O-CU selection. By optimizing globally the O-CU selection for each traffic flow, the {\NameProposal} policy consolidates traffic (Figure \ref{fig:energy_hourly_comparison}, lower off-peak energy) onto fewer active O-CUs, a significant improvement over the static 1:1 O-DU-to-O-CU mapping used by the baselines. 
    While a shortest-path algorithm provides the lowest theoretical delay, our DRL agent learns a more robust, congestion-aware routing policy. This is evident during peak hours, where the agents remain competitive on latency by intelligently routing around network bottlenecks, ensuring QoS is met under dynamic loads (see Figure \ref{fig:latency_hourly_comparison}). The choice between these algorithms presents a clear trade-off. A shortest-path baseline is straightforward for non congested networks where energy is not a concern. In contrast, our agent is superior for networks with fluctuating traffic loads and multi-objective operational goals, such as balancing performance and energy cost.
    
    It is worth noting that an actual network has a higher number of nodes/locations/links, this will increase the energy gain that we can see using a AI-driven orchestrator such as the proposed RL agent. Finally, these findings demonstrate that moving from a static, pre-calculated resource allocation model to a dynamic, AI-driven one can yield substantial benefits. Our agent's ability to learn a flexible 1:m mapping proves the value of moving beyond the rigid assumptions of current static O-RAN deployments. This capacity to learn complex, multi-objective behaviors from a reward signal represents a tangible step toward the zero-touch network automation envisioned by the O-RAN alliance, enabling more autonomous, adaptive, and energy-efficient network management.
 

\section{Conclusion}
\label{sec:Conclusion_FutureWork}

We have shown that a reinforcement learning agent augmented with a graph neural network can learn a topology-aware, multi-objective policy that jointly performs SFC routing and O-CU selection in an O-RAN setting. By enabling dynamic choice among candidate O-CU locations, the learned policy discovers energy-aware embeddings that outperform static, precomputed placements while preserving QoS guarantees. These results underscore the promise of AI-driven orchestration for realizing the flexibility envisioned by O-RAN supporting autonomous, adaptive, and energy-efficient network management under time-varying demand. 

Future work will extend this framework toward a fully multi-agent design for slice-level coordination and incorporate dynamic sleep modes for network nodes to further reduce energy consumption. 

\bibliographystyle{IEEEtran} 

\bibliography{Biblio/Placement, Biblio/Power, Biblio/GNN, Biblio/RL, Biblio/General, Biblio/BJ}
\end{document}